\documentclass[12pt,preprint]{aastex}

\shorttitle{New brown dwarf disks}
\shortauthors{Riaz \& Gizis}

\begin{document}

\title{New brown dwarf disks in the TW Hydrae Association}

\author{Basmah Riaz \& John E. Gizis}
\affil{Department of Physics and Astronomy, University of Delaware,
    Newark, DE 19716; basmah@udel.edu, gizis@udel.edu}

\begin{abstract}
In our analysis of {\it Spitzer}/IRS archival data on the stellar and sub-stellar members of the TW Hydrae Association (TWA), we have discovered two new brown dwarf disks: a flat optically thick disk around SSSPM J1102-3431 (SSSPM 1102), and a transition disk around 2MASS J1139511-315921 (2M1139). The disk structure for SSSPM 1102 is found to be very similar to the known brown dwarf disk 2MASSW J1207334-393254 (2M1207), with excess emission observed at wavelengths as short as 5 $\micron$. 2M1139 shows no excess emission shortward of $\sim$20 $\micron$, but flares up at longer wavelengths, and is the first transition disk detected among the sub-stellar members of TWA. We also report on the {\it Spitzer}/70 $\micron$ observations, and the presence of an {\it absorption} 10 $\micron$ silicate feature for 2M1207. The absorption can be attributed to a close to edge-on disk at a 75$\degr$ inclination. The 10 $\micron$ spectrum for 2M1207 shows crystalline forsterite features, with a peak in absorption near 11.3 $\micron$. No silicate absorption/emission is observed towards SSSPM 1102. While only 6 out of 25 stellar members show excess emission at these mid-infrared wavelengths, {\it all} of the TWA brown dwarfs that have been observed so far with {\it Spitzer} show signs of disks around them, resulting in a disk fraction of at least 60\%. This is a considerable fraction at a relatively older age of $\sim$10 Myr. A comparison with younger clusters indicates that by the age of the TWA ($\sim$10 Myr), the disk fraction for brown dwarfs has not decreased, whereas it drops by a factor of $\sim$2 for the higher mass stars. This suggests longer disk decay time scales for brown dwarfs compared to higher mass stars. 
\end{abstract}


\section{Introduction}
A large number of sub-stellar mass objects have been discovered over the last decade, with masses ranging from the hydrogen burning limit ($\la$ 0.075 $M_{\sun}$) down to the mass of giant planets and below the deuterium burning limit ($\la$ 0.013 $M_{\sun}$). Recent surveys in the near- and mid-infrared have confirmed the presence of disks around brown dwarfs (e.g., Muench et al. 2001; Liu et al. 2003; Apai et al. 2004; Luhman et al. 2005) and indicate a range in disk properties, similar to T Tauri disks. A detailed study of brown dwarf disks is important to understand the conditions under which planets can form in these disks. Processes that may lead to planet formation such as grain growth, crystallization and dust settling have been reported for some brown dwarfs (e.g. Apai et al. 2005), suggesting that even sub-stellar disks of a few Jupiter masses can harbor planets. It is important to determine the disk dissipation time scales for brown dwarfs, which in turn would put constraints on the time scale for the formation of planetary systems around these sub-stellar objects. Recent surveys of young ($\la$ 5 Myr) nearby star-forming regions have indicated brown dwarf disk fractions that are comparable to, or exceed, the disk fraction for higher mass stars in the same region (e.g., Liu et al. 2003; Luhman et al. 2005; Guieu et al. 2007; Bouy et al. 2007). At older ages, a disk has been detected for a $\sim$10 Myr old brown dwarf 2MASSW J1207334-393254 (2M1207) in the TW Hydrae Association (TWA) (Sterzik et al. 2004; Riaz et al. 2006). In the same association, Low et al. (2005) have determined a disk fraction of 24\% for the stellar members. The TWA is currently known to have five sub-stellar members: 2M1207, 2MASS J1139511-315921 (2M1139), SSSPM J1102-3431 (SSSPM 1102), DENIS J124514.1-442907 (DENIS 1245) and TWA 5B. 2M1207 and 2M1139 were confirmed to be TWA members by Scholz et al. (2005) based on accurate proper motions. TWA 5B is a companion to TWA 5A and lies at a projected separation of $\sim$100 AU (Lowrance et al. 1999). The accurate proper motion, photometry and spectroscopy make SSSPM 1102 very likely to be a TWA member, and it may even be a common-proper-motion companion to the star TW Hya (Scholz et al. 2005). The recently identified brown dwarf DENIS 1245 has a near-infrared spectrum that is remarkably similar to that of 2M1139 with low surface gravity features. Its position in the sky is coincident with the TWA, which makes it very likely to be a member of this association (Looper et al. 2007). The age of the TWA makes it highly favorable for the study of brown dwarf disks at relatively older ages. In this paper, we present {\it Spitzer}/IRS archival observations for three TWA brown dwarfs, along with some K- and M-dwarf members the observations for which have not been published elsewhere. We compare the disk fractions for brown dwarfs and stars at $\sim$10 Myr, and the resulting implications on the time scale for the formation of planetary systems around brown dwarfs.

\section{Observations}

We searched the {\it Spitzer} archives for IRS observations of TWA members, and were able to obtain publicly released data for 13 stars and 3 brown dwarfs. A log of the observations is given in Table 1. The IRS is composed of four modules, with two modules providing low spectral resolution (Short-Low [SL] and Long-Low [LL], $\lambda$/$\delta\lambda$ $\sim$ 90), and the other two providing high spectral resolution (Short-High [SH] and Long-High [LH], $\lambda$/$\delta\lambda$ $\sim$ 600). The SL module covers 5.2-14.5 $\micron$ in 2 orders, and the LL module covers 14-38 $\micron$, also in 2 orders. The observations in these modules are obtained at two positions along each slit. Each high-resolution (echelle) module has a single slit, with SH covering 9.9-19.6 $\micron$ and LH covering 18.7-37.2 $\micron$. The Basic Calibrated Data (BCD) were produced by the S15 pipeline at the Spitzer Science Center (SSC), which includes ramp fitting, dark sky subtraction, droop correction, linearity correction, flat fielding, and wavelength calibration. For the low-resolution spectra, the sky background was removed from each spectrum by subtracting observations taken in the same module, but with the target in the other order. We could not perform sky-subtraction for the high-resolution spectra. The spectra were extracted and calibrated using the Spitzer IRS Custom Extraction (SPICE) software provided by the SSC. The low resolution spectra were extracted using a variable-width column extraction that scales with the width of the wavelength-dependent point spread function. The SH and LH spectra were extracted with a full-slit extraction. The low-resolution spectra were calibrated with HR 7341 (K1 III), and the high-resolution ones with HR 6688 (K2 III). The spectra from each module were then averaged to obtain a single spectrum for that module. These were merged together to obtain the full 5.2-38 $\micron$ spectrum. The signal-to-noise ratio (S/N) for the spectra varies strongly with wavelength, and the noise levels are found to be higher at the red end of each order. The S/N of these spectra ranges from $>$100 for the K- and M-dwarfs to a few for the brown dwarfs.

Along with the archival data, we present MIPS/70 $\micron$ observations for 2M1207. These were obtained as part of the General Observer (GO) program (PID 40922). We used an exposure time of 10s and 20 template cycles. Aperture photometry was performed on the artifact-corrected mosaic images using the task PHOT under the IRAF package APPHOT. We used an aperture radius of 5 pixels, a background annulus of 6-12 pixels, a zero-point flux of 0.775 Jy, and an aperture correction of 2.07. The magnitude error is found to be $\sim$0.4 mag. The calibration uncertainty in the MIPS bands is $\sim$10\%.

\clearpage
\begin{deluxetable}{cccccc}
\tabletypesize{\scriptsize}
\tablecaption{Observing Log}
\tablewidth{0pt}
\tablehead{
\colhead{Name}  & \colhead{Spectral Type\tablenotemark{a}} & \colhead{Observing Mode} & \colhead{{\it Spitzer} ID} 
} 
\startdata
TW Hya	&	K8e	&	SL, LL, Mapping	&	2	\\
TWA 2AB	&	M2e+M2	&	SL, LL, Staring	&	2	\\
Hen 3-600	&	M3e+M3.5	&	SL (Mapping), SH, LH, Staring	&	2 \& 30300	\\
HD 98800AB	&	K4+K5	&	SL (Mapping), SH, LH, Staring	&	2 \& 124	\\
TWA 5A	&	M3e	&	SL, LL, Staring	&	2	\\
TWA 6	&	K7	&	SL, LL, Staring	&	2	\\
TWA 7	&	M1	&	SL, LL, Staring	&	2	\\
TWA 8A	&	M2e	&	SL, LL, Staring	&	2	\\
TWA 8B	&	M5	&	SL, LL, Staring	&	2	\\
TWA 9AB	&	K5+M1	&	SL, LL, Staring	&	2	\\
TWA 10	&	M2.5	&	SL, LL, Staring	&	2	\\
HR 4796A	&	A0	&	SL, SH, LH, Mapping	&	2	\\
TWA 15AB	&	M1.5+M2	&	SL, LL, Staring	&	20691	\\
TWA 16	&	M1.5	&	SL, LL, Staring	&	20691	\\
TWA 20	&	M2	&	SL, LL, Staring	&	20691	\\
TWA 22	&	M5	&	SL, LL, Staring	&	20691	\\
TWA 24	&	K3	&	SL, LL, Staring	&	20691	\\
2M1207	&	M8	&	SL, LL, Staring	&	30540	\\
2M1139	&	M8	&	SL, LL, Staring	&	2	\\
SSSPM 1102	&	M8.5	&	SL, LL, Staring	&	30540	\\
\enddata
\tablenotetext{a}{Spectral types for all stars are from Low et al. (2005), except for 2M1207 and 2M1139 from Gizis (2002), and SSSPM 1102 from Scholz et al. (2005).}

\end{deluxetable}
\clearpage

\section{The IRS Spectra}

Figure 1a shows the spectra for the four strong disks, TW Hya, Hen 3-600, HD 98800B and HR 4796A, all of which were first detected by {\it IRAS} and are known to exhibit strong excess emission at mid- and far-infrared wavelengths. We have included in Fig. 1 the ground-based 10 $\micron$ photometry measured by Jayawardhana et al. (1999), the {\it Spitzer} 24 $\micron$ photometry measured by Low et al. (2005), and the {\it IRAS} 12 and 25 $\micron$ photometry. The photometric measurements by Jayawardhana et al. and Low et al. are found to be in good agreement with the IRS spectra, although the {\it IRAS} 12 $\micron$ flux densities are stronger than the values derived from the IRS spectra for TW Hya and HR 4796A. Since the narrowest IRS slit is 3.6$\arcsec$ wide, the IRS observations include emission from all components for the Hen 3-600, HD 98800 and HR 4796 systems. The spectra for all four of these indicate the presence of strong disks, with little or no excess emission detected shortward of $\sim$8 $\micron$. The SEDs of the T Tauri stars TW Hya and Hen 3-600 show no hint of turning over, even out to 160 $\micron$ (Low et al. 2005). For the HD 98800 system, Low et al. (2005) have shown that the B-component generates the huge infrared excess that is well-fitted by a single blackbody with {\it T} = 160-170 K, whereas component A lacks any measurable excess. Likewise the infrared excess emission for the A0 V star HR 4796A can be fitted by a single blackbody at a cooler temperature of 108 K (Low et al. 2005). The stars TW Hya, Hen 3-600 and HD 98800B also exhibit broad 10 $\micron$ silicate emission features (discussed further in $\S$\ref{silicate}). The spectra for some of the other K- and M-type TWA members are shown in Figures 1 b-d. These are arranged in an order of increasing spectral type. The main feature in the spectra for these low-mass stars at wavelengths of 5-40 $\micron$ is the $H_{2}O$ absorption band near 6.3 $\micron$ (Cushing et al. 2006). The strength in this band is found to increase with increasing spectral type, until it saturates in the T spectral class. We have measured the depth of $H_{2}O$ absorption for TWA K-and M-dwarfs using the IRS-$H_{2}O$ index defined by Cushing et al. as:

\begin{equation}
IRS-H_{2}O = f_{6.25}/(0.562 f_{5.80} + 0.474 f_{6.75}),
\end{equation}

\noindent where $f_{\lambda}$ is the mean flux density in a 0.15 $\micron$ window centered around $\lambda$. The errors in the indices were computed from the uncertainties in the mean flux densities, and were found to be between 0.001 and 0.002. Larger values of this index indicate stronger $H_{2}O$ absorption. Fig. 1e shows a plot of the IRS-$H_{2}O$ indices versus the spectral type for the TWA K- and M-dwarfs. We note that we have only 11 data points, and we did not include 2M1207 and SSSPM 1102 as the observed flux densities for those around 6 $\micron$ are dominated by the disk emission. As can be seen, the strength in the $H_{2}O$ absorption decreases from spectral type K5 to M0, and then increases again towards later types. Two exceptions are TWA 8A (M2) and 8B (M5) that show stronger $H_{2}O$ absorption than the brown dwarf 2M1139 (M8). The spectra shown in Figs. 1 b-d are all found to be photospheric, other than TWA 7 which is known to harbor a cold debris disk (Low et al. 2005). The spectrum for this M1 dwarf is photospheric for $\lambda$ $\la$ 24 $\micron$, but the flux densities begin to rise at longer wavelengths and are found to be in excess of the estimated photospheric emission. The spectrum for TWA 8B (M5) in Fig. 1d shows slight emission in the 10 $\micron$ silicate feature. The flux density derived from the spectrum is 14$\pm$0.4 mJy, compared to 10$\pm$2 mJy reported by Jayawardhana et al. (1999). This small excess could be due to the presence of an optically thin disk, however Low et al. (2005) have not reported excess emission in the MIPS bands for this mid-M dwarf. The separation of $\sim$1$\arcmin$ between TWA 8A and 8B is large enough to be resolved by the IRS slits, and thus there should not be any contribution from the A-component to the slight excess emission observed for the B-component. A similar modest excess was measured by Jayawardhana et al. (1999) for TWA 5A. However, the spectrum for this M3 dwarf has a similar slope as the M2.5 star TWA 10 (Fig. 1d), and there is no excess emission seen near 10 $\micron$. As noted by Jayawardhana et al. (1999), if a more conservative limit of {\it K} - {\it N} $\sim$ 0.3 is used to define an excess, then the observed emission at 10 $\micron$ for TWA 5A would be photospheric. To summarize, among the stellar members of TWA, we do not find evidence of any significant excess emission at wavelengths of 5-40 $\micron$ for any of the stars other than the six already known disk-bearing members mentioned above.

\clearpage
\begin{figure}
 \begin{center}
    \begin{tabular}{ccc}      
      \resizebox{70mm}{!}{\includegraphics[angle=0]{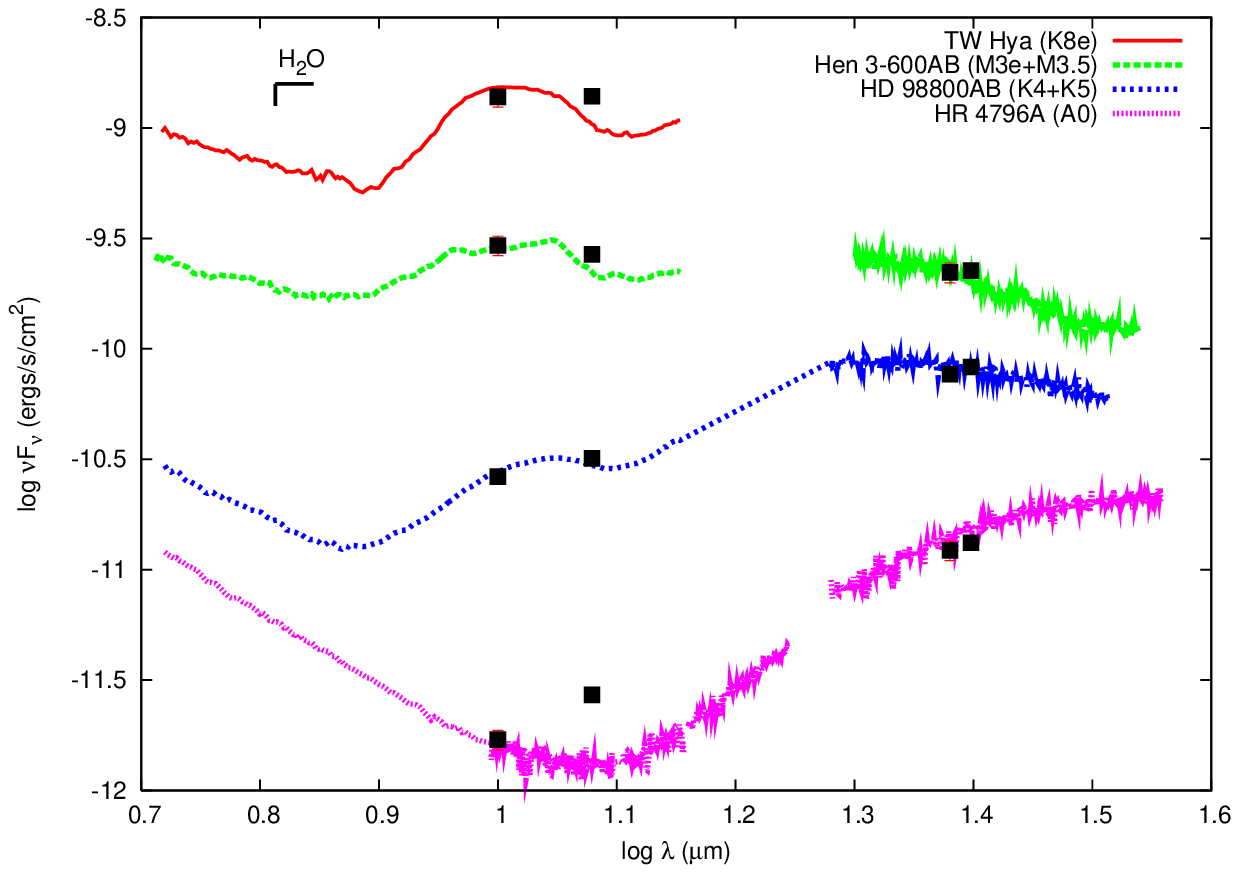}} &
      \resizebox{70mm}{!}{\includegraphics[angle=0]{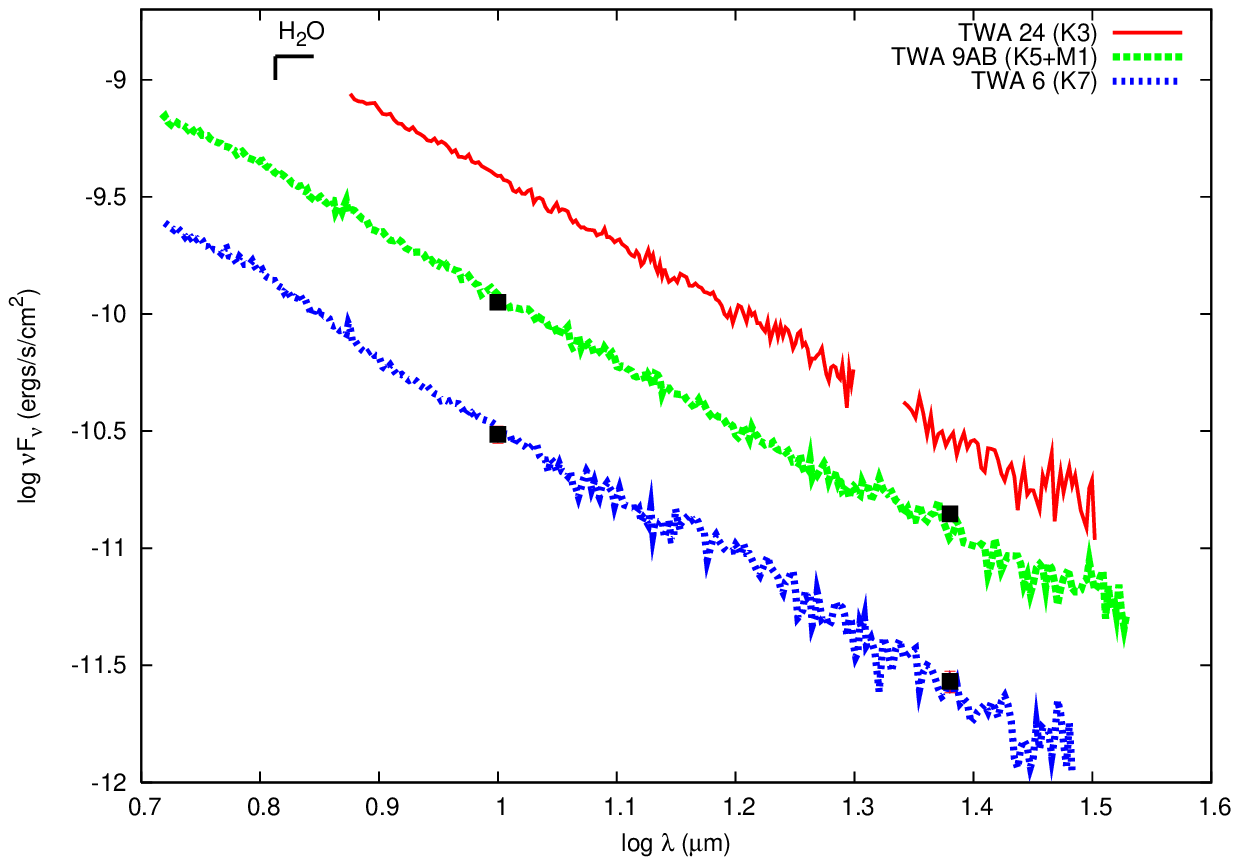}} \\
      \resizebox{70mm}{!}{\includegraphics[angle=0]{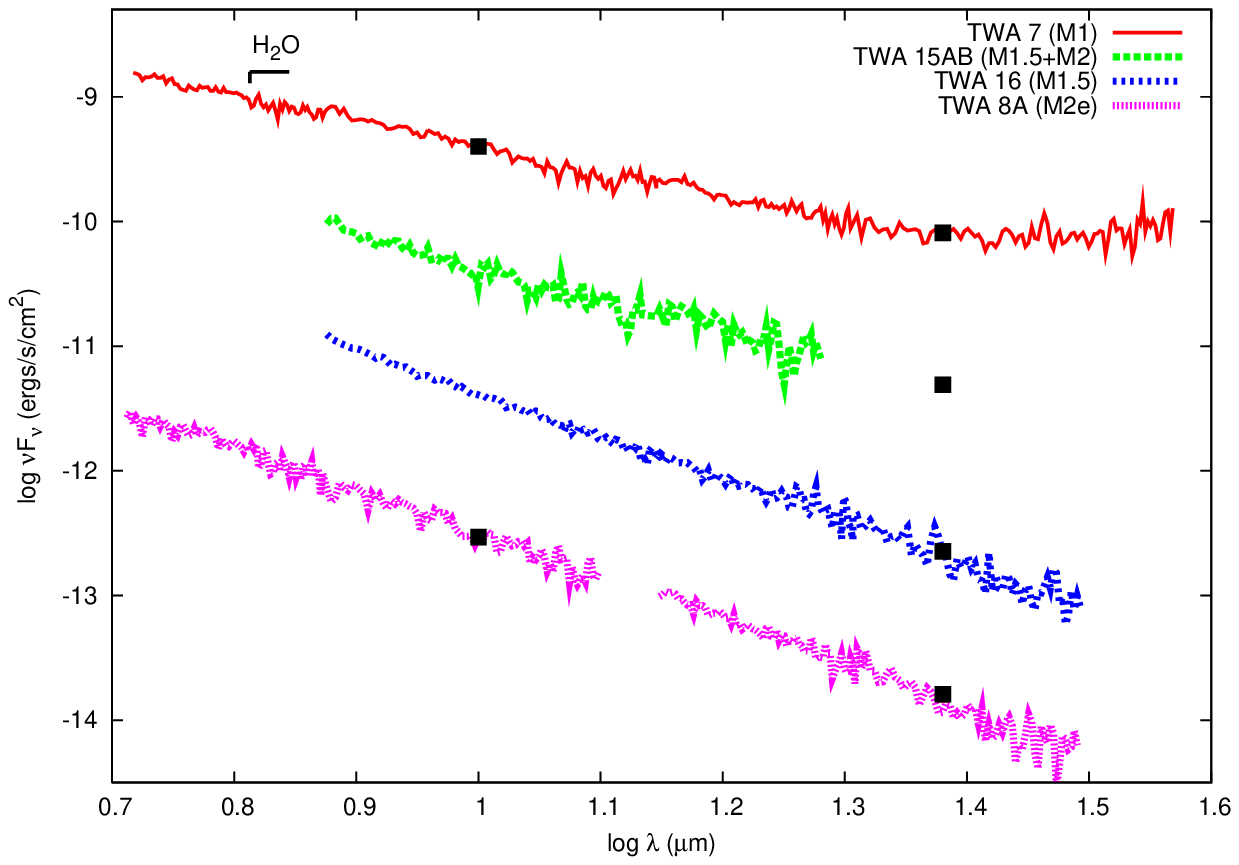}} &
      \resizebox{70mm}{!}{\includegraphics[angle=0]{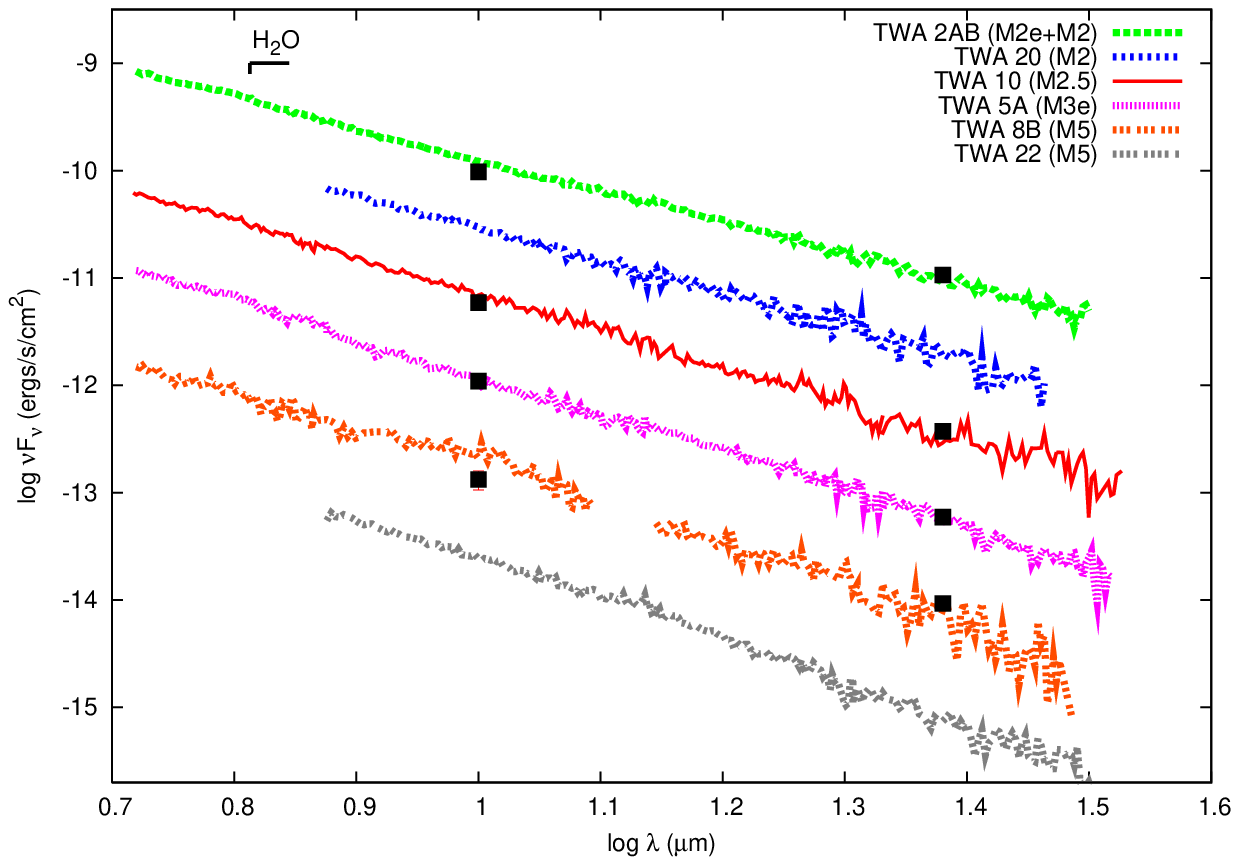}} \\
      \resizebox{70mm}{!}{\includegraphics[angle=0]{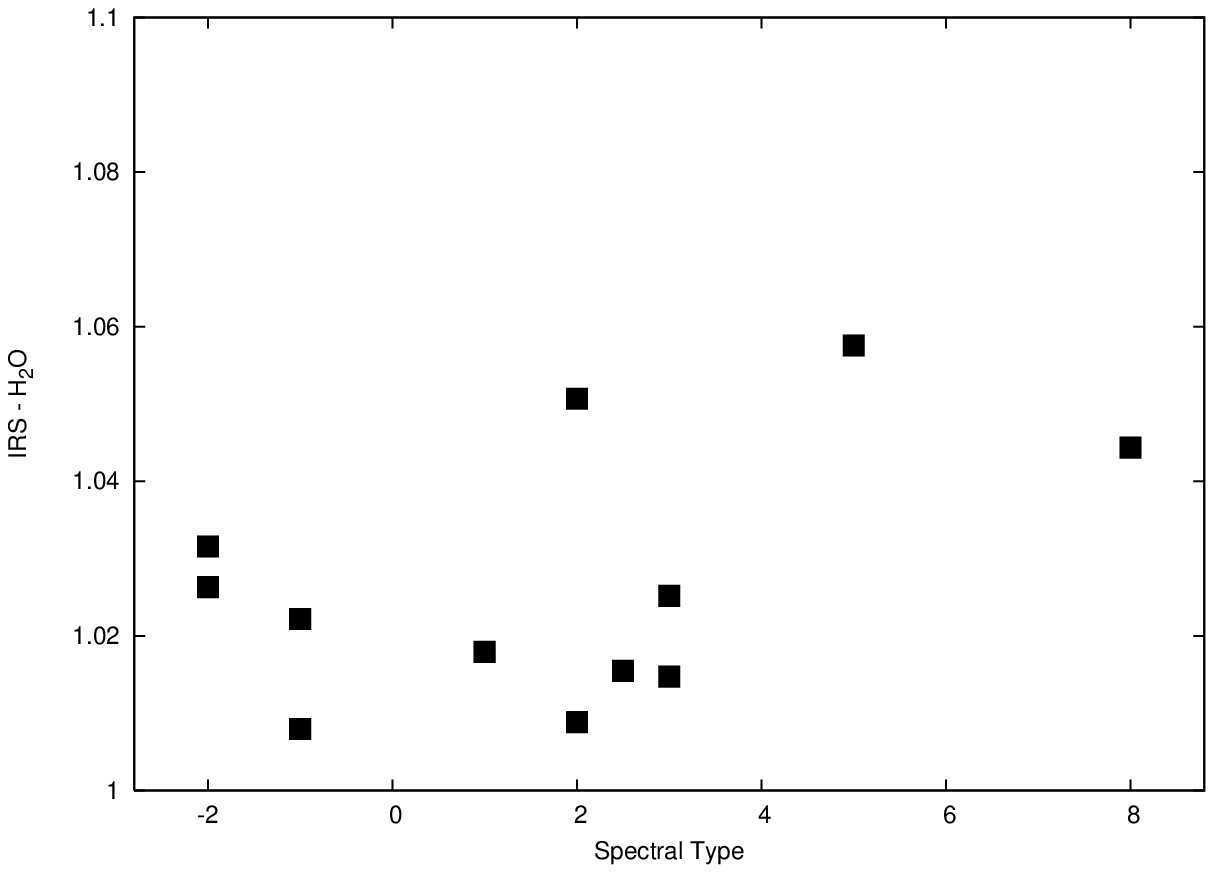}} \\      
    \end{tabular}
    \caption{IRS spectra for the TWA stellar members. Black filled squares denote 10 and 24 $\micron$ measurements from Jayawardhana et al. (1999) and Low et al. (2005), respectively. {\it Bottom}-Strength in the $H_{2}O$ absorption band versus the spectral type.}
  \end{center}
\end{figure}
\clearpage

We now look at the IRS spectra for the TWA brown dwarfs 2M1207, SSSPM 1102 and 2M1139 (Fig. 2a). Also included in this figure are the IRS spectra for the field dwarfs LHS 3003 (M7), vB 10 (M8) and BRI 0021-0214 (M9). In comparison with these, the flat spectra for SSSPM 1102 and 2M1207 clearly indicate the presence of excess emission at the IRS wavelengths of 5-40 $\micron$. The brown dwarf 2M1207 is known to harbor a circumsub-stellar disk of gas and dust (Sterzik et al. 2004; Riaz et al. 2006). The spectra for 2M1207 and SSSPM 1102 are found to be very similar, other than the 10 $\micron$ silicate feature which is in absorption for 2M1207, but absent for SSSPM 1102. The full 5-40 $\micron$ emission seen in the IRS observation for SSSPM 1102 is found to be in excess of the estimated photospheric emission at these wavelengths, as can be seen from Fig. 2b, where we have used the NextGen model (Hauschildt et al. 1999) for a $T_{eff}$ of 2600 K and log {\it g} = 3.5 to fit the atmosphere spectrum of the central sub-stellar source. The spectrum for 2M1139 is found to be photospheric for wavelengths shorter than $\sim$20 $\micron$, but a rise in the flux densities is observed at longer wavelengths. Riaz et al. (2006) had reported an excess emission in the MIPS/24 $\micron$ band for 2M1139 that is $\sim$3 times brighter than the predicted photospheric emission. They had discussed the possibility for this brown dwarf disk to flare up at 70 $\micron$ in a fashion similar to the more massive M1 V star TWA 7, for which the observed flux level at this wavelength is $\sim$40 times brighter than the estimated photospheric emission (Low et al. 2005). Although the LL mode observation for 2M1139 has a low S/N ratio ($\sim$9), the excess emission at 24 $\micron$ and a rise in the observed flux densities at longer wavelengths strongly indicate the presence of a transition disk around it (Fig. 2c), and is the first such detection among the sub-stellar members of TWA. We will be obtaining MIPS/70 $\micron$ observations (PID 40922) for 2M1139, which will allow us to further confirm the presence of a disk around it. To summarize, {\it all} of the TWA brown dwarfs that have been observed so far with {\it Spitzer} are found to have disks around them, compared to just 6 out of the 25 stellar members. 

\clearpage
\begin{figure}
 \begin{center}
    \begin{tabular}{ccc}      
      \resizebox{70mm}{!}{\includegraphics[angle=0]{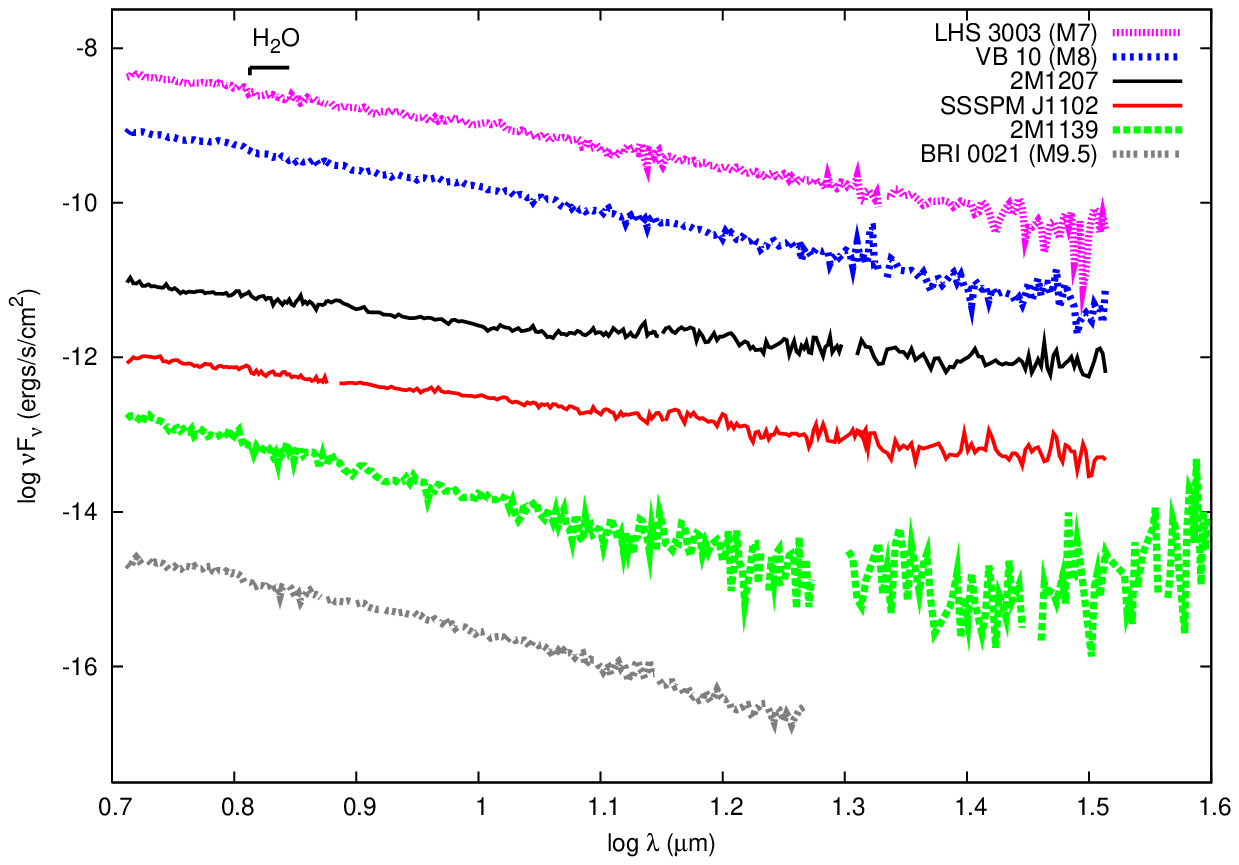}} \\
      \resizebox{70mm}{!}{\includegraphics[angle=0]{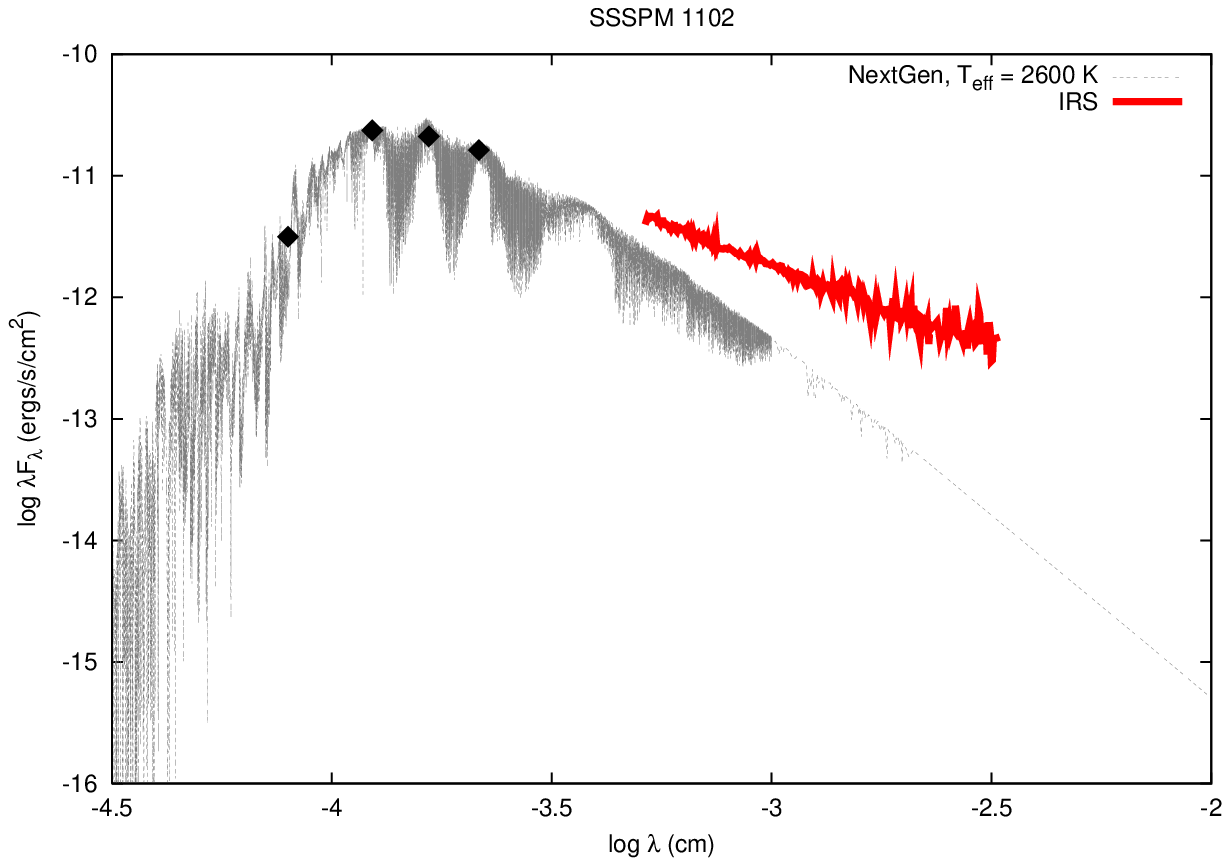}} \\
      \resizebox{70mm}{!}{\includegraphics[angle=0]{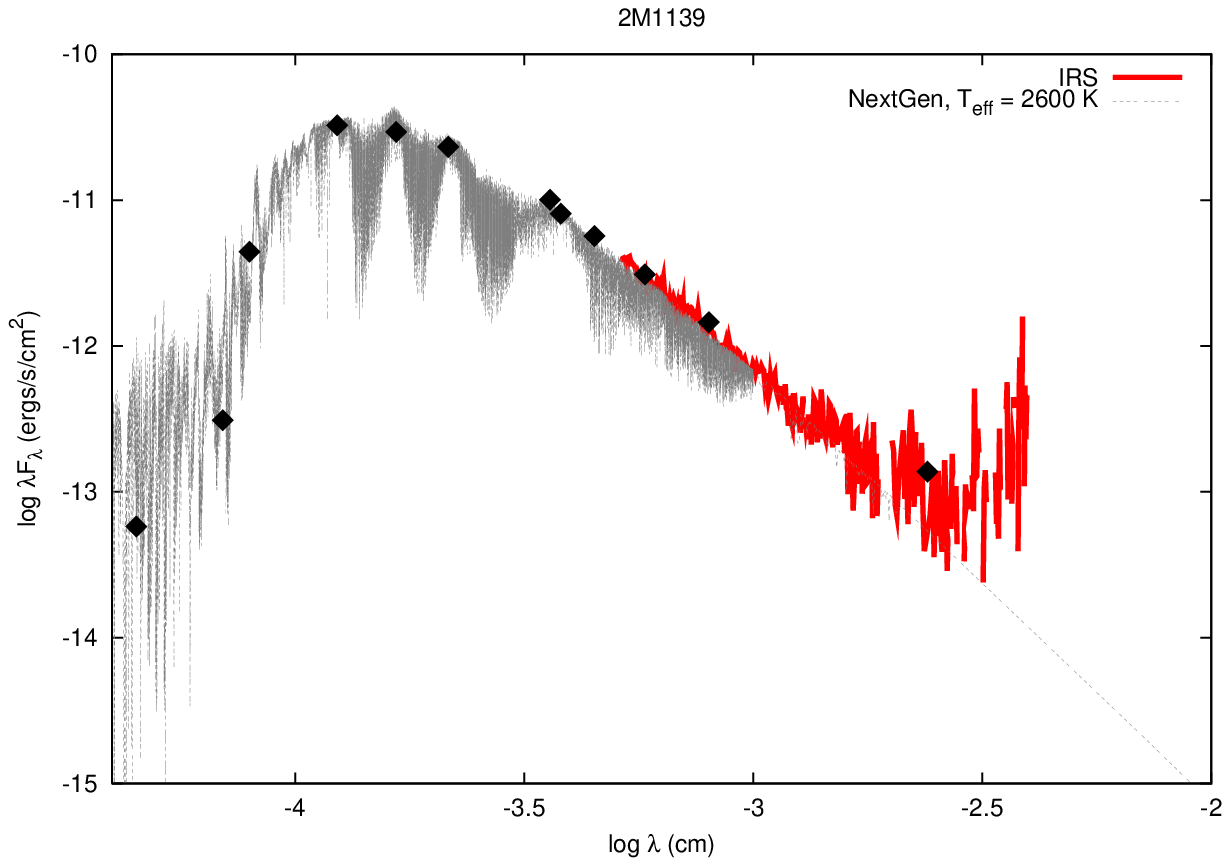}} \\     
    \end{tabular}
    \caption{{\it Top}-IRS spectra for TWA brown dwarfs. Included for comparison are spectra for the field dwarfs LHS 3003 (M7), vB 10 (M8) and BRI 0021-0214 (M9). SEDs for SSSPM 1102 ({\it middle}) and 2M1139 ({\it bottom}). The grey dotted line in these SEDs denotes the NextGen model for a $T_{eff}$ of 2600 K and log {\it g} = 3.5. Black filled diamonds denote {\it VRI} magnitudes from SIMBAD, {\it JHK} from 2MASS, and IRAC and MIPS from Riaz et al. (2006).}
  \end{center}
\end{figure}
\clearpage

In Table 2, we have compiled the accretion, activity and mid-infrared measurements for the nine disk-bearing objects in TWA. The 10\% width is the full width of the H$\alpha$ line at 10\% of the peak, and is a strong accretion indicator. Objects with 10\% width $>$ 200 km/s are considered to be accretors (Jayawardhana et al. 2003). The three disk candidates in TWA that are also accretors (TW Hya, Hen 3-600 and 2M1207) have {\it v}sin{\it i} $\la$13 km/s (Table 2), while the non-accreting objects with disks as well as the diskless stars show a range in {\it v}sin{\it i} with values $<$5 km/s up to 25 km/s (Jayawardhana et al. 2006). This is consistent with the disk-locking scenario in which stars with disks rotate slowly due to rotational braking (e.g. Jayawardhana et al. 2006). Fig. 3a indicates a strong correlation between disk and accretion. The fractional disk luminosities $L_{IR}/L_{*}$ increase with increasing accretion rates. This is expected as the stronger accretors would have more circumstellar material, and thus can reprocess a larger fraction of the stellar luminosity into the infrared. The two brown dwarf disks have $L_{IR}/L_{*}$ intermediate between the two cold debris disks (TWA 7 and 13) and the strong disks. Low et al. (2005) had reported a bimodal distribution of warm dust in the TWA, based on the observed excess emission at 24 $\micron$. Including the sub-stellar disks makes the bimodal nature of 24 $\micron$ excess less pronounced. No connection between accretion and X-ray activity is evident among the TWA stars (Fig. 3b). The two non-accretors with disks (TWA 7 and 13), as well as the diskless objects show similar X-ray emission as the accreting disks. Given that TW Hya and Hen 3-600 are the only two accretors, it is difficult to determine any dependence of accretion on activity for these stars. Among the TWA brown dwarfs, both 2M1207 and SSSPM 1102 have been found to be sub-luminous in X-rays (Gizis \& Bharat 2004; Stelzer et al. 2007). In comparison, TWA 5B is a non-accretor (10\% width = 162 km/s; Mohanty et al. 2003), no IR excess has yet been reported for this object, but it shows strong X-ray emission ($L_{X}$ = 4 $10^{27}$ ergs/s; Tsuboi et al. 2003). There may thus be some relation between accretion and activity among the sub-stellar objects.

\clearpage
\begin{deluxetable}{ccccccc}
\tabletypesize{\scriptsize}
\tablecaption{The TWA disks}
\tablewidth{0pt}
\tablehead{
\colhead{Object} & \colhead{\.{M} ($M_{\sun}$/yr)}  & \colhead{$L_{X}$ (ergs/s)} & \colhead{{\it v}sin{\it i}} & \colhead{$L_{IR}/L_{*}$} & \colhead{10\% width (km/s)} & \colhead{References\tablenotemark{a}} 
} 
\startdata
TW Hya & 4 $10^{-10}$ & 2.6 $10^{30}$ & 10.6 & 2.7 $10^{-1}$ & 425 & 1,2,3,2,3 \\
Hen 3-600 & 5 $10^{-11}$ & 8.5 $10^{29}$ & 11.6 & 1.2 $10^{-1}$ & 262 & 1,2,3,2,3 \\
HD 98800B\tablenotemark{b} & $<$$10^{-11}$ & 1.5 $10^{30}$ & - & 2.2 $10^{-1}$ & - & -,2,-,2,- \\
HR 4796A\tablenotemark{b} & $<$$10^{-11}$ & -& -& 4.8 $10^{-3}$ & -& -,-,-,2,- \\
TWA 7\tablenotemark{b} & $<$$10^{-11}$ & 9.6 $10^{29}$ & $<$ 5.0 & 2.0 $10^{-3}$ & 109 & -,2,3,2,3 \\
TWA 13A\tablenotemark{b} & $<$$10^{-11}$ & 1.3 $10^{30}$ & 10.5 & 8.6 $10^{-4}$ & 129 & -,2,3,2,3 \\
2M1207 & $10^{-10.1...-9.8}$ & $<$1.2 $10^{26}$ & 13 & 6.0 $10^{-2}$ & 170...320 & 4,5,6,7,6\&4 \\
SSSPM 1102 & $<10^{-11}$ & $<$5.3 $10^{26}$ & - & 5.0 $10^{-2}$ & 194 & 8,4,-,7,8 \\
2M1139\tablenotemark{b} & $<$$10^{-11}$ & - & 25 & - & 111 & -,-,6,-,6 \\
\enddata
\tablenotetext{a}{(1) Muzerolle et al. (2000), (2) Low et al. (2005), (3) Jayawardhana et al. (2006), (4) Stelzer et al. (2007), (5) Gizis \& Bharat (2004), (6) Mohanty et al. (2003), (7) this work, (8) Scholz et al. (2005).}
\tablenotetext{b}{These objects are non-accreting, so the accretion rates can be expected to be lower than $10^{-11} M_{\sun}$/yr.}
\end{deluxetable}
\clearpage

\begin{figure}
 \begin{center}
    \begin{tabular}{ccc}      
      \resizebox{90mm}{!}{\includegraphics[angle=0]{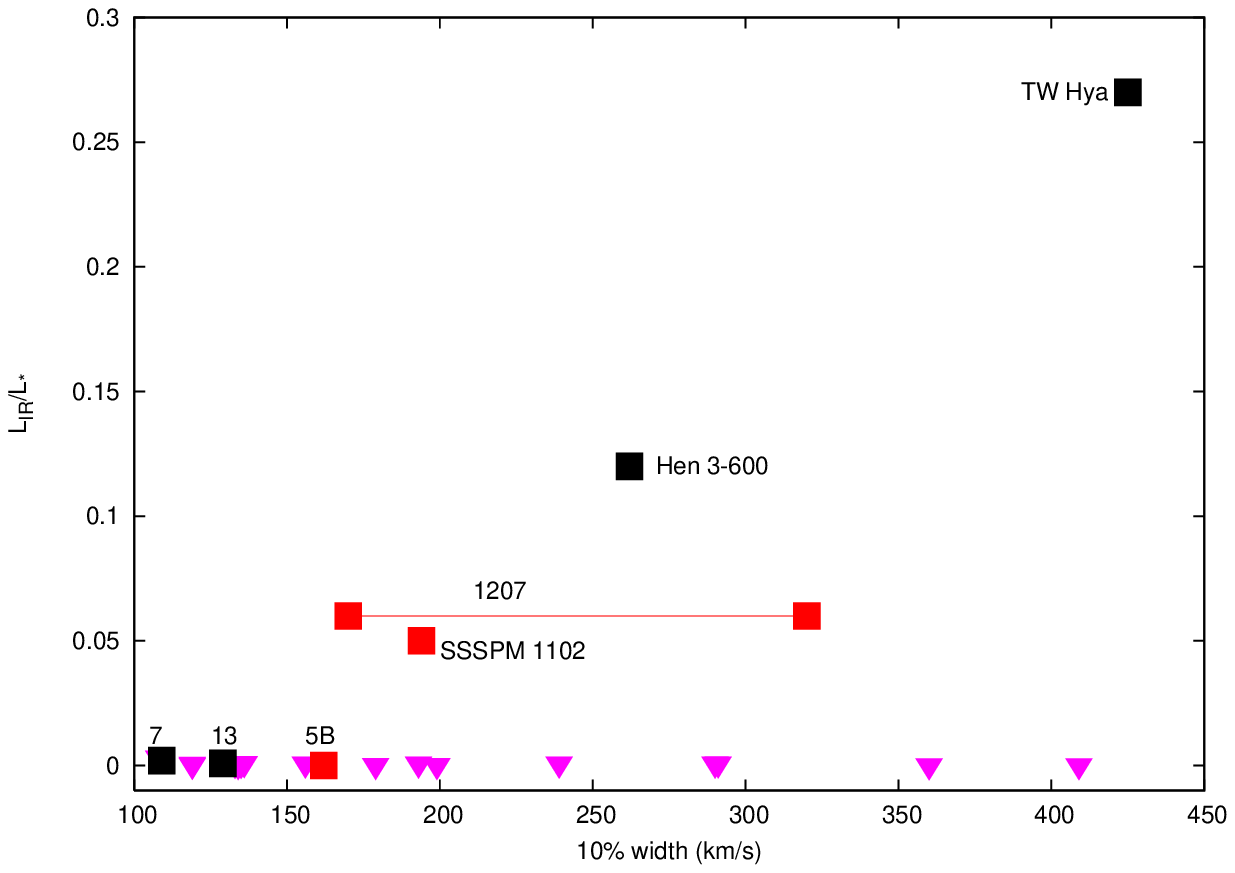}} \\
      \resizebox{90mm}{!}{\includegraphics[angle=0]{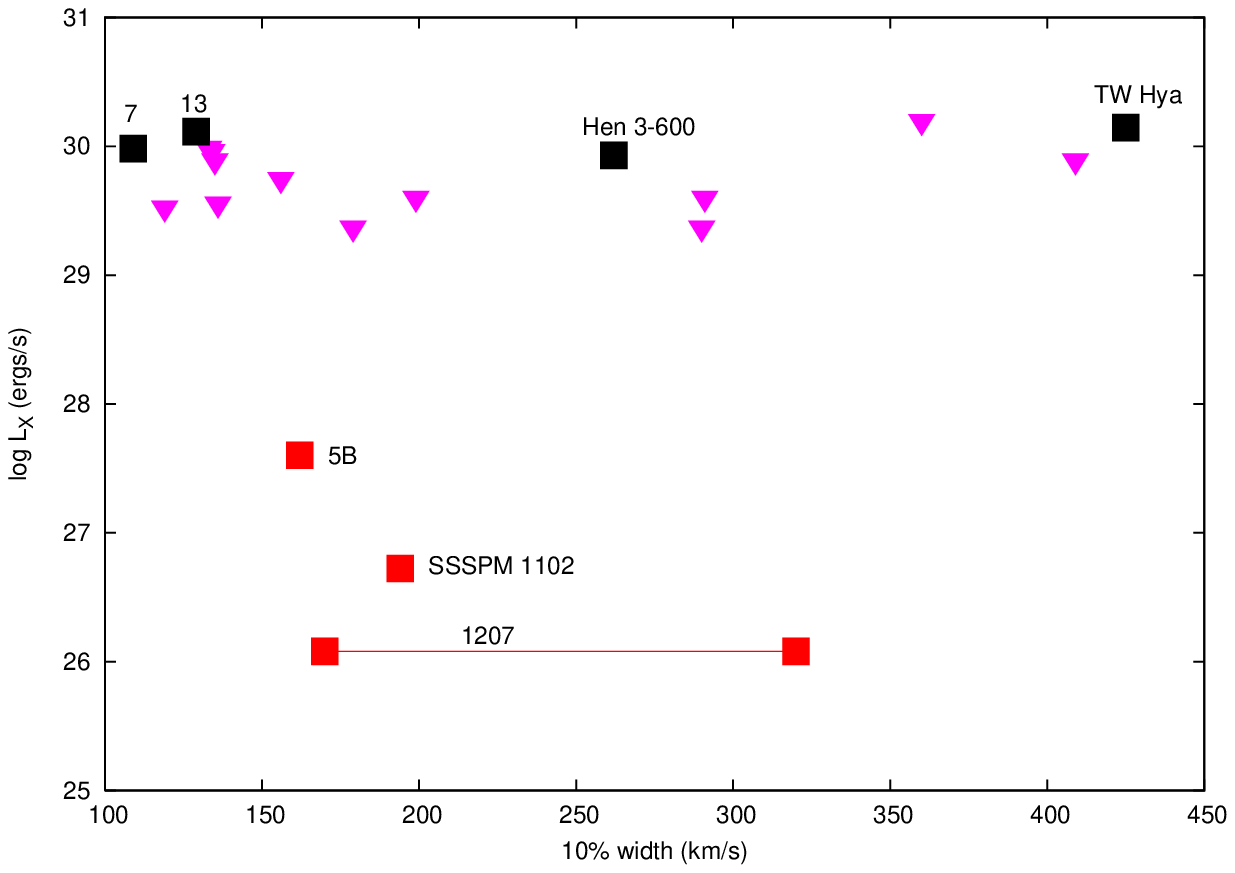}} \\  
    \end{tabular}
    \caption{Relations between disk diagnostics, accretion and activity. The 10\% widths are from Jayawardhana et al. (2006); values for $L_{IR}/L_{*}$ and $L_{X}$ are from Low et al. (2005). Black solid squares denote the stellar members with disks, red solid squares denote the brown dwarfs. Solid triangles denote the non-disk bearing members; the fractional disk luminosities for these objects have been determined from the MIPS observational upper limits. {\it Top} (a)- Accretion versus fractional disk luminosity. {\it Bottom} (b)- Accretion versus X-ray luminosity. }
  \end{center}
\end{figure}
\clearpage

\section{The 10 $\micron$ Silicate Feature}
\label{silicate}

The IRS spectra cover the 10$\mu$m region of the silicate emission/absorption feature, and can be used to determine the degree of growth and crystallinity of the silicate grains present in the disk. Fig. 4 shows the 8-13$\mu$m continuum-subtracted spectra for the 5 disks in TWA that display excess emission at these wavelengths. The continuum was obtained by connecting the two end points of the 10 $\micron$ spectrum. TW Hya, HD 98800B and Hen 3-600 are known to exhibit broad emission features. Recent detailed modeling by Sargent et al. (2006) indicates very little crystalline material (only $\sim$1\% by mass) for TW Hya but a substantial content ($\sim$24\% by mass) of large amorphous grains. In comparison, Hen 3-600 shows a mixture of crystalline and large silicates ($\sim$36\% and $\sim$32\% by mass, respectively) (Sargent et al. 2006). Detailed decomposition of the broad silicate feature for HD 98800B by Sh\"{u}tz et al. (2005) indicates the presence of highly processed dust, dominated by both large amorphous olivine and crystalline forsterite. 

The brown dwarf 2M1207 shows a weak absorption silicate feature. Kessler-Silacci et al. (2005; hereafter K05) have discussed that the strength in the silicate absorption is a function of the optical depth, while the peak wavelength is dependent on the dust composition. The IRS spectra for Class I objects in their sample exhibit smooth, featureless absorption profiles with a strong, narrow peak at 9.6$\micron$, indicative of amorphous silicates. Also, stronger absorption is observed in the more embedded protostars in their sample. In comparison with the protostars, 2M1207 exhibits shallower absorption, as expected from a Class II source. Absorption in the silicate feature for 2M1207 can be explained by the presence of a highly inclined disk, which would result in an increase in the optical depth of the absorbing dust and give the same effect as being a more embedded star. A mixture of crystalline forsterite and large amorphous pyroxene can account for the observed feature in 2M1207. It shows a broad flat profile that is indicative of the presence of large ($>$2 $\micron$) grains. The observed peaks near $\sim$11.3 and $\sim$10.2 $\micron$ can be attributed to crystalline forsterite. The smaller peak observed at $\sim$9.5 $\micron$ is indicative of amorphous pyroxene. There is also a small peak at $\sim$10.7 $\micron$ that can be owed to crystalline orthoenstatite (Honda et al. 2003). No silicate absorption/emission is observed towards SSSPM 1102. A possible explanation for the lack of 10 $\micron$ feature could be grain growth to sizes larger than the wavelength at which they radiate ($>$10 $\micron$). Such large grains would then only produce a blackbody continuum (Bouwman et al. 2001). Flat spectra such as these could also indicate a dearth of small ($<$2 $\micron$) grains in the upper disk layers probed by the 10 $\micron$ feature. We do not find any evidence for a 20 $\micron$ silicate feature in 2M1207 or SSSPM 1102. The low S/N ($\sim$9) in the LL module for these very faint sources makes it difficult to identify any significant absorption/emission due to silicate at 20 $\micron$. The two brown dwarfs show weaker, flatter features than the TWA stars, which can be explained by the differences in the location of the silicate formation zone. This is defined to be the radial zone where the cumulative flux reaches 20\%-80\% of the total 10 $\micron$ emission, and varies as $R_{10}$ = 0.35 AU $(L_{*}/L_{\sun})^{0.56}$ (Kessler-Silacci et al. 2007; hereafter K07). The silicate formation thus occurs at smaller radii ($R_{10} <$ 0.001-0.1 AU) in disks around brown dwarfs than in disks around T Tauri stars ($R_{10} >$ 0.1-3 AU), and would therefore be more affected by inner disk evolution due to processes such as grain growth and/or dust crystallization (K07). Furthermore, the size of the region producing the optically thin silicate feature is smaller for late-type stars, resulting in weaker features (Sicilia-Aguilar et al. 2007).

\begin{figure}
 \begin{center}
    \begin{tabular}{cc}      
      \resizebox{60mm}{!}{\includegraphics[angle=0]{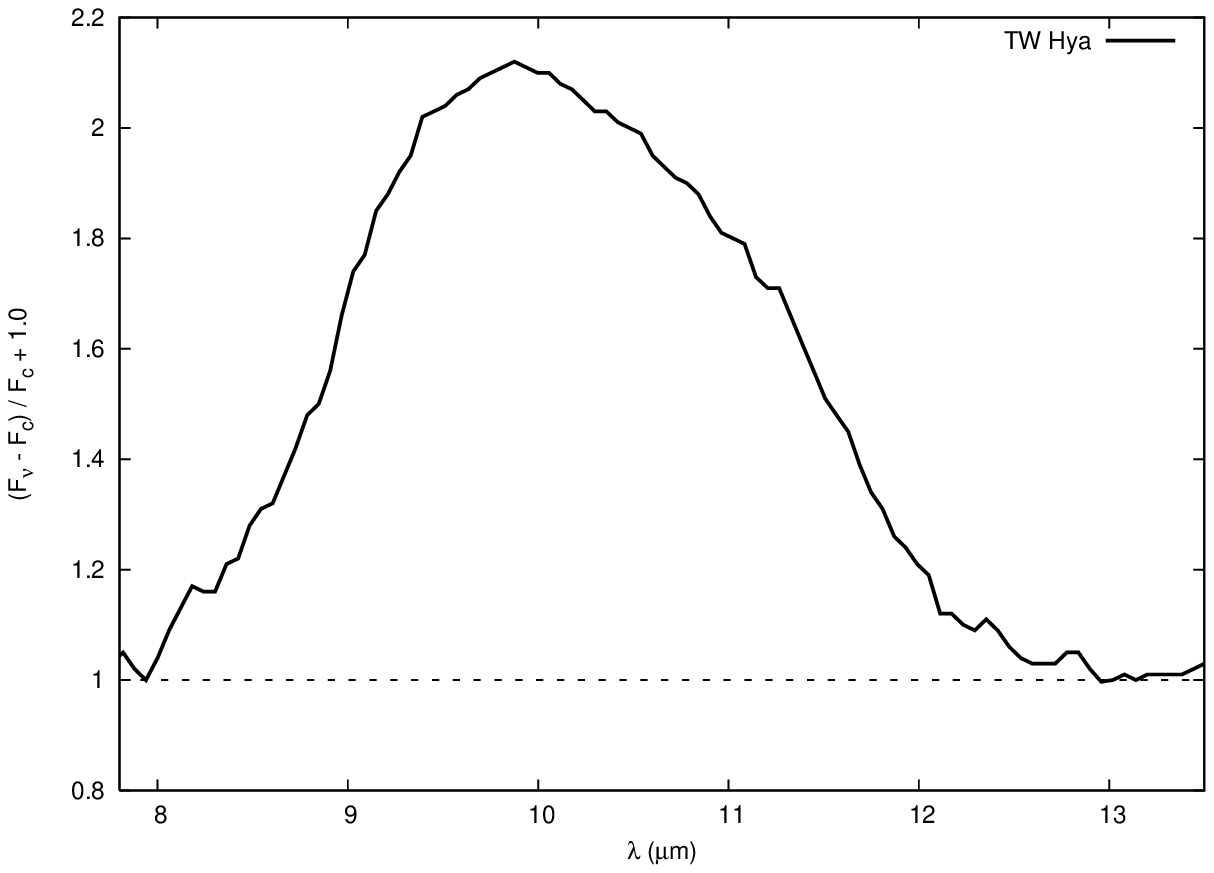}} \\
      \resizebox{60mm}{!}{\includegraphics[angle=0]{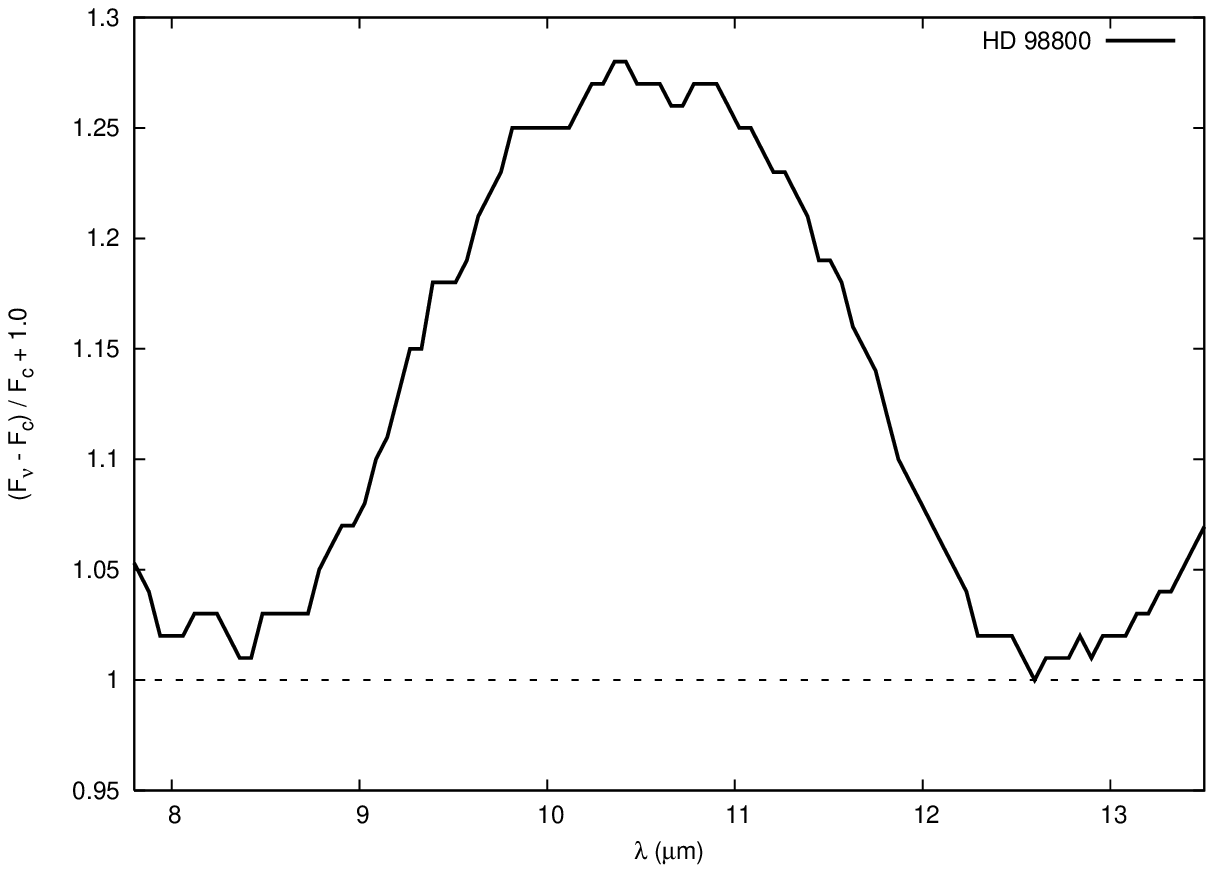}} \\
      \resizebox{60mm}{!}{\includegraphics[angle=0]{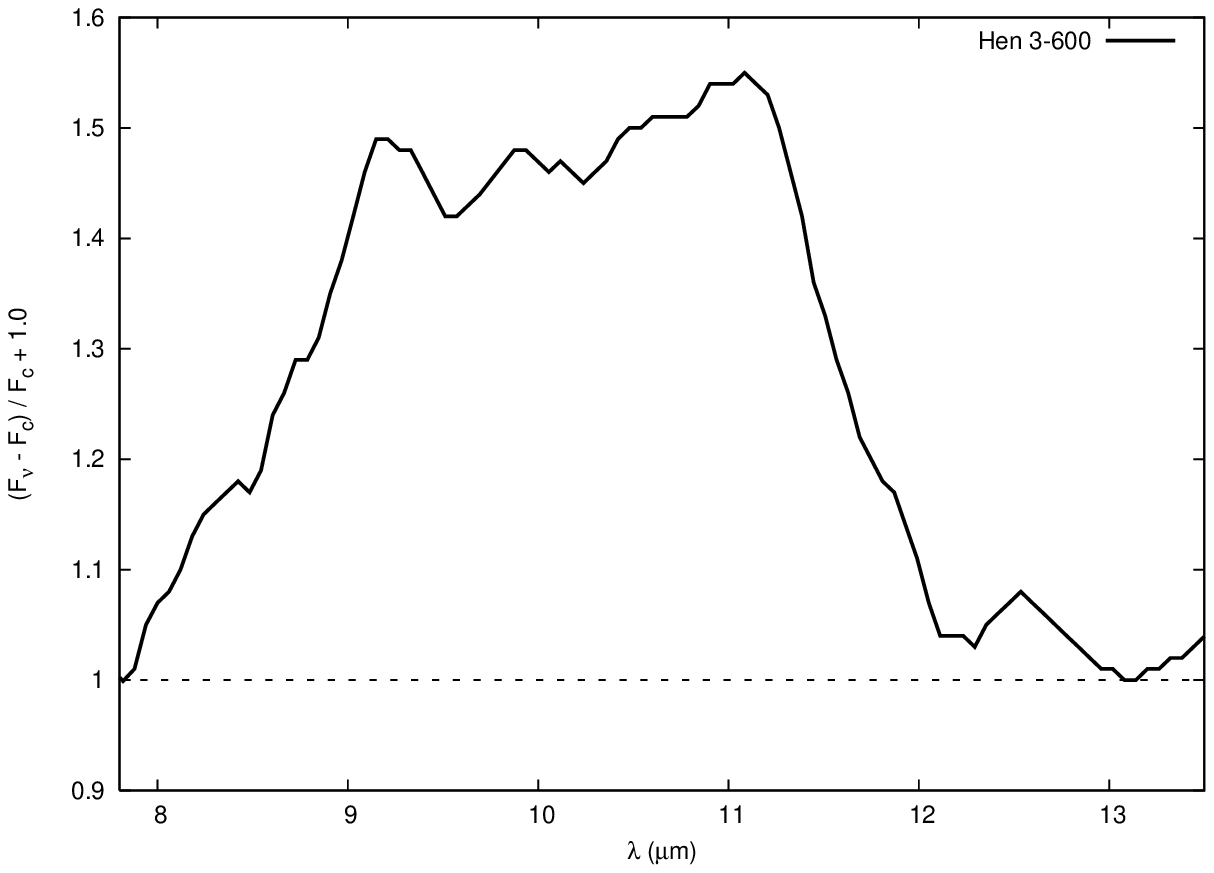}} \\ 
      \resizebox{60mm}{!}{\includegraphics[angle=0]{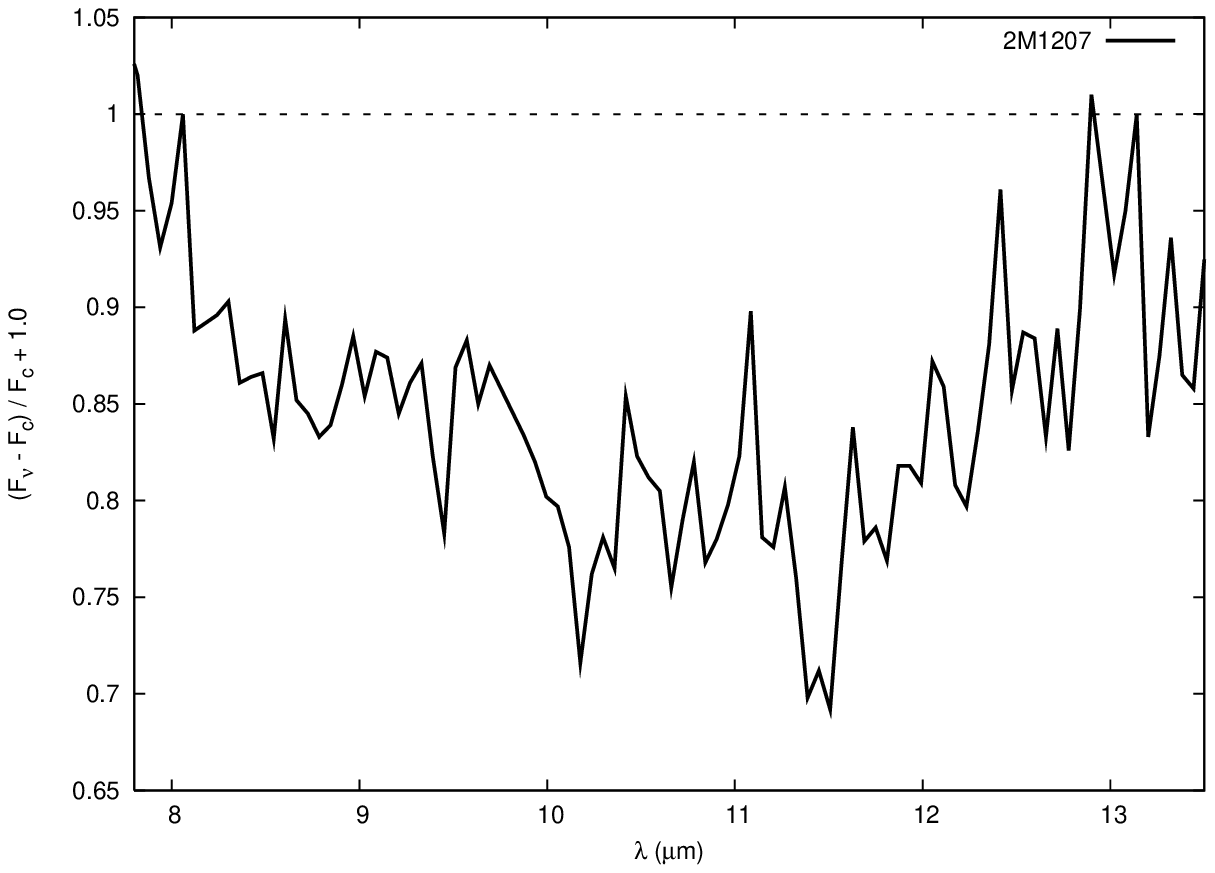}} \\
      \resizebox{60mm}{!}{\includegraphics[angle=0]{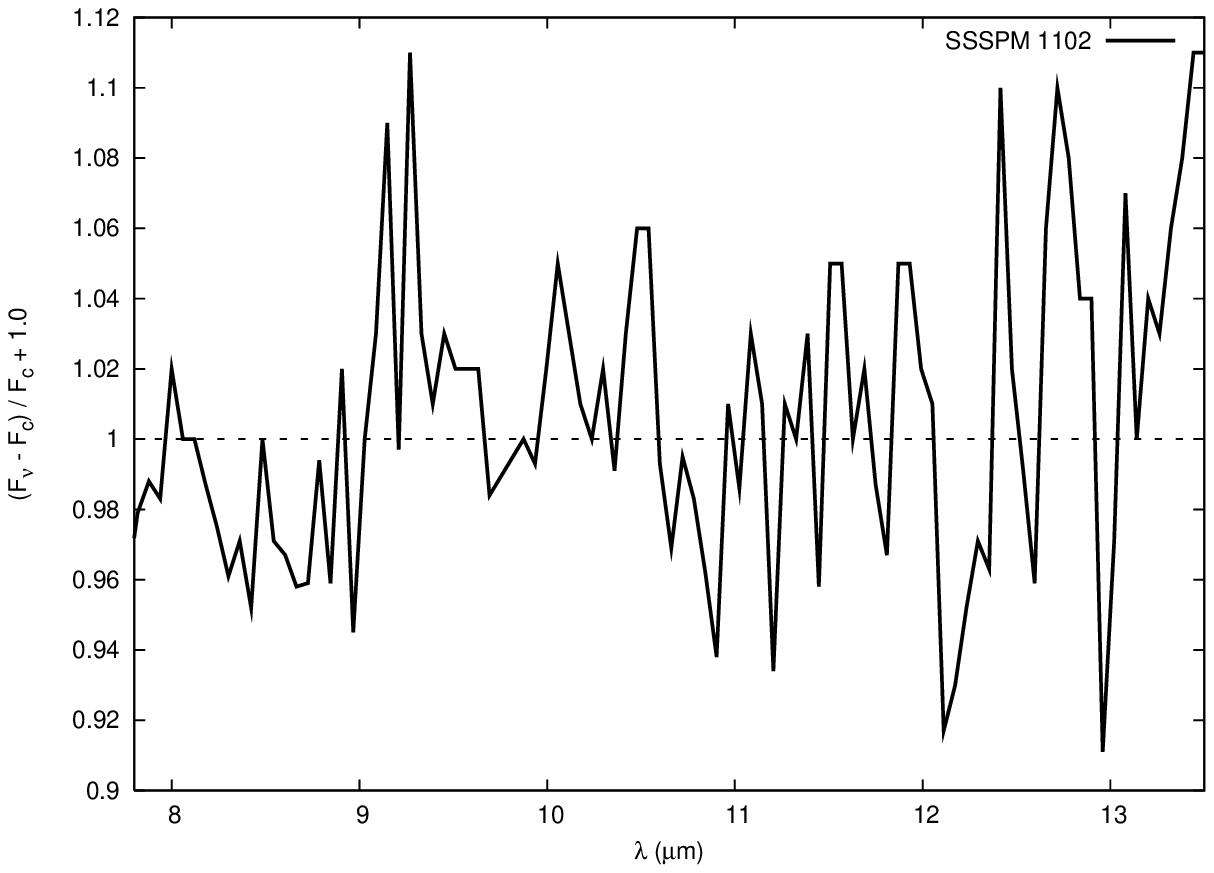}} \\     
    \end{tabular}
    \caption{The normalized 10 $\micron$ spectra in units of ($F_{\nu}-F_{c})/F_{c}$ + 1 for TWA stars and brown dwarfs. The dashed line represents the continuum level. From top to bottom: TW Hya, HD 98800B, Hen 3-600, 2M1207 and SSSPM 1102. }
  \end{center}
\end{figure}

Fig. 5 shows a plot of the "shape" vs. "strength" of the 10 $\micron$ silicate feature for the TWA stars and brown dwarfs. The shape is indicated by the 11.3 to 9.8 $\micron$ flux ratio, while the feature strength can be estimated by the peak flux above/below the continuum (e.g., K05). Table 3 lists the values for these parameters. The vertical dashed line in Fig. 5 at $F_{peak}$ = 1 represents the continuum; objects to the right of this line show emission in the silicate feature, while those to the left show absorption. TW Hya, Hen 3-600 and HD 98800B follow the trend shown by the solid line, that indicates a change in the feature shape from peaked to flat to be caused by an increase in the amorphous grain sizes (K07). Hen 3-600 and HD 98800B, with broad features due to the presence of a mixture of amorphous and crystalline silicates, lie above and to the left of TW Hya that shows a comparatively narrower profile with a peak indicative of amorphous olivine. The brown dwarf SSSPM 1102 that lacks a silicate feature lies at the top left near $F_{peak}$$\sim$1. Among the absorption features, a smaller value for $F_{11.3}/F_{9.8}$ indicates more absorption near 11.3 $\micron$, while the feature strength decreases with increasing $F_{peak}$. The absorption feature thus becomes flatter in moving from top left to the bottom right, which could be explained by an increase in the amorphous grain sizes. For features showing stronger absorption near 11.3 $\micron$ ($F_{11.3}/F_{9.8}$ $<$ 1), crystallization may become more dominant and grain growth alone may not be able to explain the flattening for such features. In their study of silicate emission features in young brown dwarf disks, Apai et al. (2005) have shown that for highly processed grains, the feature strength may increase with increasing crystallinity and will reverse the observed correlation. A similar reversal may also be true in the case of the absorption features. The location of 2M1207 below and to the right of the protostars is indicative of highly processed grains, due to grain growth and/or dust crystallization.

\begin{deluxetable}{ccccccc}
\tabletypesize{\scriptsize}
\tablecaption{Spectral parameters for the silicate feature}
\tablewidth{0pt}
\tablehead{
\colhead{Name}  & \colhead{$\lambda_{peak}$ ($\micron$)} & \colhead{$F_{peak}$} & \colhead{$F_{11.3}$/$F_{9.8}$} & \colhead{$L_{*}$/$L_{\sun}$} 
} 
\startdata
TW Hya & 9.87 & 2.12 &  0.78 & 0.24 \\
HD 98800B & 10.35 & 1.28 & 0.97 & 0.53 \\
Hen 3-600 & 11.08 & 1.55 & 0.98 & 0.35 \\
2M1207 & 11.50 & 0.69 & 0.91 & 0.0031 \\
SSSPM 1102 & 9.26 & 1.11 & 1.0 & 0.0031 \\
\enddata
\end{deluxetable}

\begin{figure}
 \begin{center}
    \begin{tabular}{ccc}      
      \resizebox{130mm}{!}{\includegraphics[angle=0]{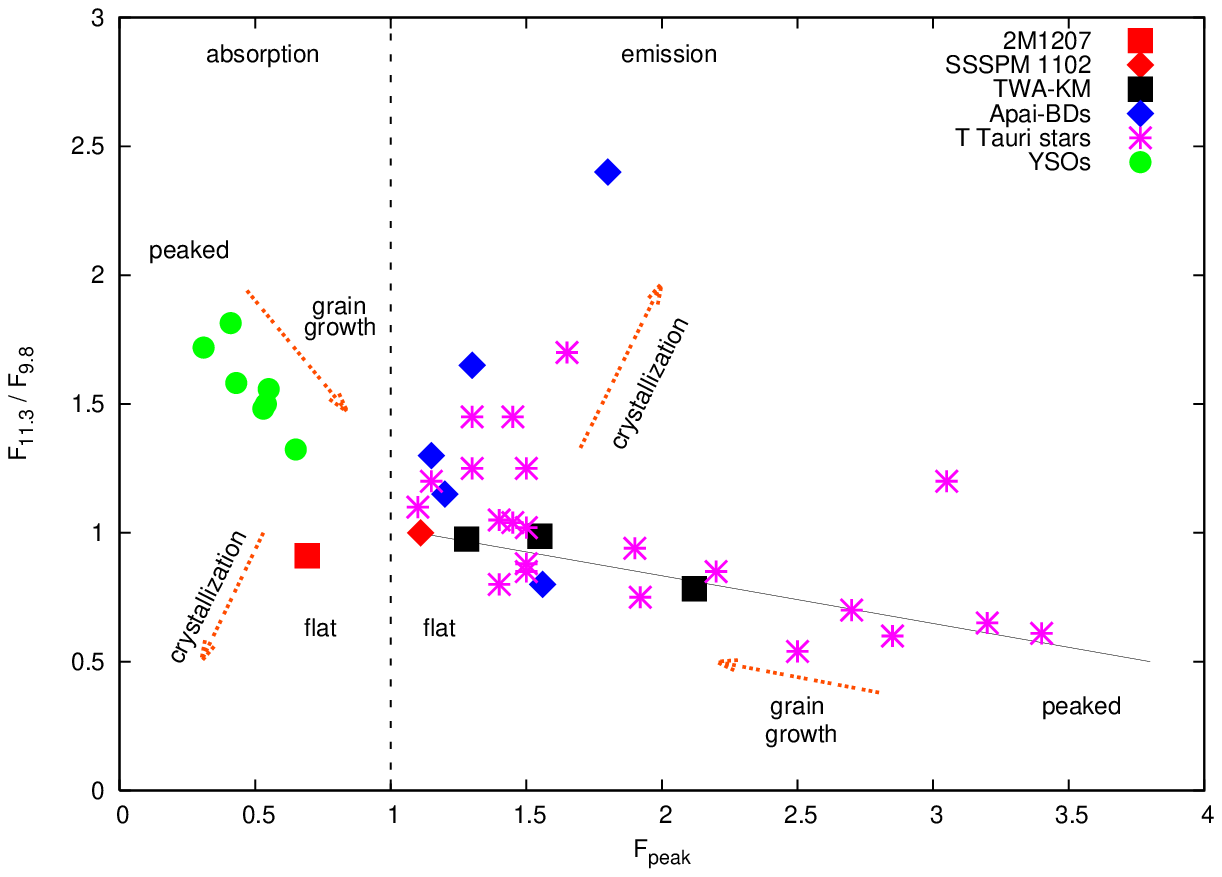}} \\
    \end{tabular}
    \caption{Shape versus strength of the silicate feature. 2M1207 and SSSPM 1102 are represented by red square and triangle, respectively. The three TWA stars TW Hya, Hen 3-600 and HD 98800B are denoted by black squares. The vertical dashed line represents the continuum: objects to the left show absorption in the feature while those to the right of this line show emission. Solid line denotes the fit obtained by K07 to the points with y $<$ 1.1, y = -(0.18$\pm$0.02)x + (1.23$\pm$0.03). Also included for comparison are young (1-3 Myr) brown dwarfs in the Chamaeleon I star-forming region (Apai et al. 2005; blue triangles), T Tauri stars from Przygodda et al. (2003) and K07 (denoted by crosses), and young low-mass embedded sources from K05 (denoted by green circles). }
  \end{center}
\end{figure}

\section{Disk Modeling}
\label{model}

Figure 6a shows the extent of excess emission at 24 $\micron$ for the TWA stars and brown dwarfs. The dot-dashed line represents the limit of $F_{24}/F_{K}$ under the assumption that both bands lie on the Rayleigh-Jeans tail of the stellar spectrum. The four strong disks are about a factor of 100 brighter at 24 $\micron$ relative to the {\it K}-band than other stars in the TWA (Low et al. 2005). The brown dwarf disks lie intermediate between the two debris disks TWA 7 and 13, and the four strong disks. The $F_{24}$-to-$F_{K}$ ratio for 2M1139 is higher than TWA 7 or TWA 13, indicating a warmer debris disk around this brown dwarf comapred to the cold 80 K disk around TWA 7 or the 65 K disk around TWA 13AB. The dashed line in Fig. 6a represents a geometrically thin, optically thick flat disk with a spectral slope $\lambda F_{\lambda} \propto \lambda^{-4/3}$. The three transition disks with no observed excess emission shortward of $\sim$20 $\micron$ lie below this line. 2M1207 and SSSPM 1102 lie just at the dashed line, indicating optically thick flat disks around these two brown dwarfs.

\begin{figure}
 \begin{center}
    \begin{tabular}{ccc}      
      \resizebox{80mm}{!}{\includegraphics[angle=270]{f6a.epsi}} \\
      \resizebox{80mm}{!}{\includegraphics[angle=0]{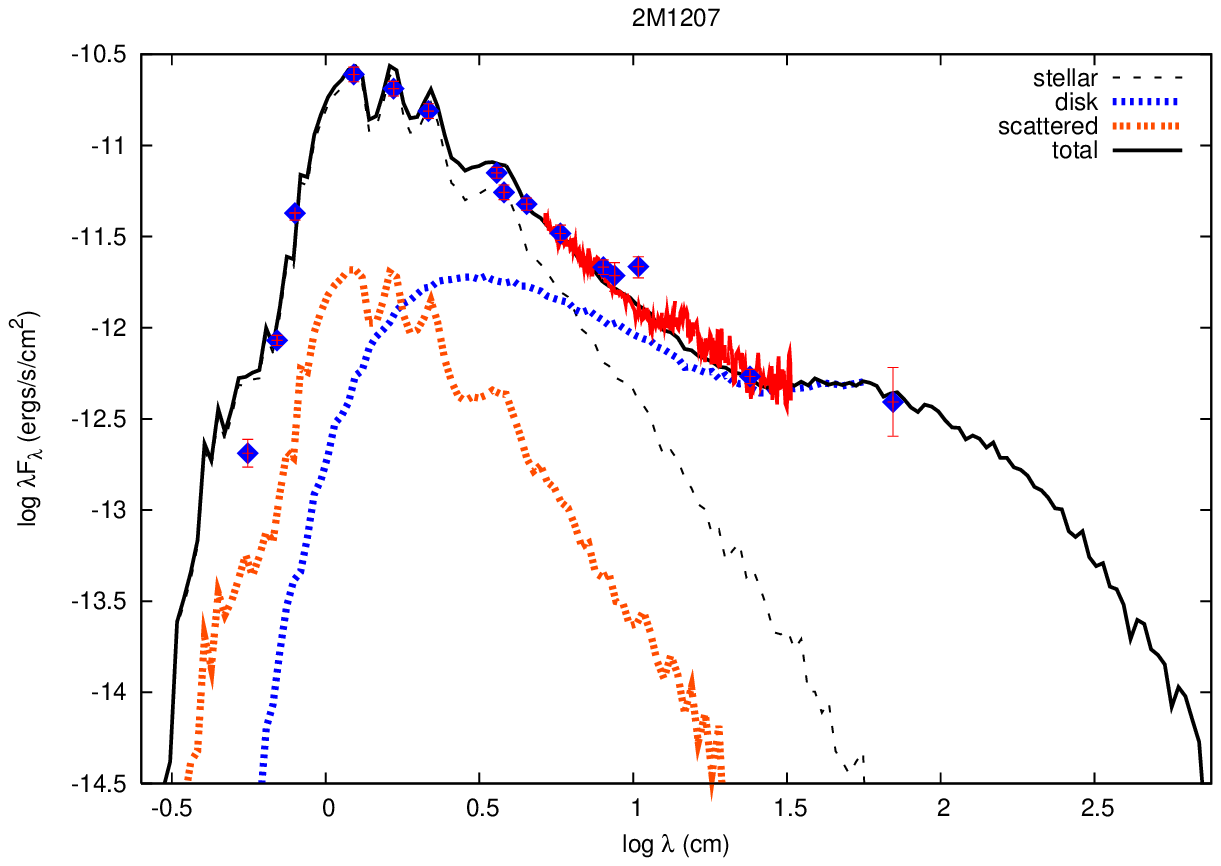}} \\
      \resizebox{80mm}{!}{\includegraphics[angle=0]{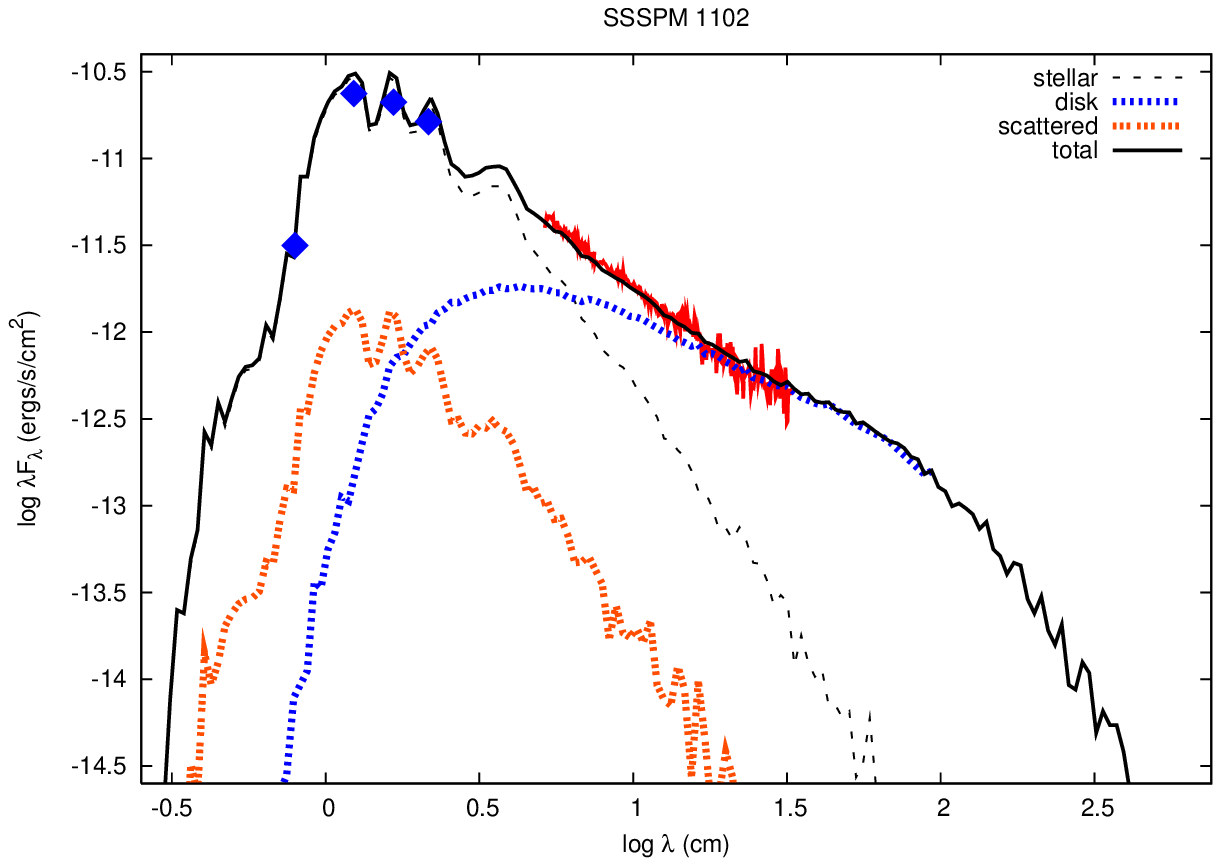}} \\     
    \end{tabular}
    \caption{{\it Top}- The extent of excess emission at 24 $\micron$ for the TWA stars and brown dwarfs. The dot-dashed line represents the limit of $F_{24}/F_{K}$ under the assumption that both bands lie on the Rayleigh-Jeans tail of the stellar spectrum. The dashed line represents a geometrically thin, optically thick flat disk with a spectral slope $\lambda F_{\lambda} \propto \lambda^{-4/3}$. Best-fit disk models for 2M1207 ({\it middle}) and SSSPM 1102 ({\it bottom}). Separate contributions from the disk, the stellar photosphere and the scattered flux are indicated. The {\it VRI} magnitudes for 2M1207 are from Gizis et al. (2007), {\it JHK} from 2MASS, IRAC and MIPS/24 $\micron$ from Riaz et al. (2006), and MIPS/70 $\micron$ from this work.}
  \end{center}
\end{figure}

We have used the 2-D radiative transfer code by Whitney et al. (2003) to model the disks around 2M1207 and SSSPM 1102. The circumstellar geometry consists of a rotationally flattened infalling envelope, bipolar cavities, and a flared accretion disk in hydrostatic equilibrium. The disk density is proportional to $\varpi^{-\alpha}$, where $\varpi$ is the radial coordinate in the disk midplane, and $\alpha$ is the radial density exponent. The disk scale height increases with radius, $h=h_{0}(\varpi / R_{*})^{\beta}$, where $h_{0}$ is the scale height at $R_{*}$ and $\beta$ is the flaring power. Since we are fitting a disk source, the envelope was turned off by setting its mass infall rate equal to zero. For the stellar parameters, we have used $T_{eff}$=2550K and $M_{*}$=0.024 $M_{\sun}$ (Gizis 2002). Using this $T_{eff}$ and $M_{*}$, the evolutionary tracks by Burrows et al. (1997) imply $R_{*} \sim$ 0.26 $R_{\sun}$. The NextGen (Hauschildt et al. 1999) atmosphere file for a $T_{eff}$ of 2600K, and log {\it g} = 3.5 was used to fit the atmosphere spectrum of the central sub-stellar source. A distance of 54$\pm$3 pc (Gizis et al. 2007) was used to scale the output fluxes from the models to the luminosity and distance of the two brown dwarfs. 

Figs. 6b-c show the model fits obtained for 2M1207 and SSSPM 1102. Also indicated are the separate contributions from the disk, the stellar photosphere and the scattered flux. In the disk midplane, we have used large grains with a size distribution that decays exponentially for sizes larger than 50 $\micron$ up to 1 mm. In the disk atmosphere, we have used grains of sizes with $a_{max}\sim$ 1 $\micron$. Decreasing the grain size in the disk midplane results in more emission in the silicate feature, while the effect is opposite in the disk atmosphere. We have explored a range of inclination angles to the line of sight. The silicate feature is found to be more in absorption for highly inclined disks. Due to binning of photons in the models, there are a total of 10 viewing angles, with edge-on covering 87-90$\degr$ inclinations. For 2M1207, a more edge-on viewing angle of 75$\degr$ provides a good fit to the observed silicate absorption, while a 63$\degr$ inclination is a good fit to the flat feature for SSSPM 1102. On the basis of the redshifted absorption component in the H$\alpha$ profile, Scholz et al. (2005) had concluded that 2M1207 is seen with an inclination of {\it i} $\ga$ 60$\degr$. Our model fits are thus consistent with previous conclusion on the inclination angle. For SSSPM 1102, a disk mass of $10^{-5} M_{\sun}$ and flaring power $\beta$ = 1.02 provide a good fit to the disk emission. For 2M1207, there is a change in slope observed between 24 and 70 $\micron$ due to a slight increase in the scale height at 70 $\micron$. Increasing the disk mass to $10^{-4} M_{\sun}$, and $\beta$ to 1.06 fits both the IRS spectrum and the 70 $\micron$ point. The mass accretion rate was set at $10^{-10} M_{\sun}/yr$ for 2M1207 and $10^{-11} M_{\sun}/yr$ for SSSPM 1102 (Scholz et al. 2005; Stelzer et al. 2007). The disk outer radius was set to 100 AU. Since the disk mass was fixed, smaller $R_{out}$ resulted in larger optical depth. The changes however are more evident at far-IR/sub-mm wavelengths. We varied the inner disk radius $R_{in}$ in multiples of $R_{sub}$, which is the dust sublimation radius and varies with the stellar radius and temperature, $R_{sub} = R_{*} (T_{sub}/T_{*})^{-2.085}$ (Whitney et al. 2003). $T_{sub}$ is the dust sublimation temperature and was set to 1600K. For 2M1207 and SSSPM 1102, 1$R_{sub} \sim$ 3$R_{*}$. Increasing the inner disk radius results in higher fluxes near the 10 $\micron$ silicate band and at longer wavelengths. The best-fit (based on a reduced $\chi^{2}$ comparison) was obtained using $R_{in}$=1$R_{sub}$, which implies an absence of an inner hole in the disk since it would have to be larger than the dust sublimation radius. The fractional disk luminosities for 2M1207 and SSSPM 1102 are found to be 0.06 and 0.05, respectively, and lie intermediate between the strong disks such as TW Hya with $L_{IR}/L_{*}$ of 0.27 and the cool debris disks such as TWA 7 with $L_{IR}/L_{*}$ of 0.002 (Low et al. 2005). We note here that while our IRAC and MIPS photometry matches the IRS spectrum for 2M1207 (Fig. 6b), the flux density reported by Sterzik et al. (2004) at 10.4 $\micron$ is higher by a factor of $\sim$1.6 than that derived from the spectrum. Riaz \& Gizis (2007) had found a good fit to Sterzik et al. measurement using sub-micron sized grains in the upper disk layers, which results in a narrow and peaked 10 $\micron$ silicate emission feature. The IRS spectrum, on the other hand, shows absorption  and thus rules out the measurement reported by Sterzik et al. (2004).

\section{Disk Lifetimes}
\label{fractions}

It seems like {\it all} of the brown dwarfs in the TWA have disks around them, as shown here for the 3 out of 5 sub-stellar members observed with {\it Spitzer}. We note that we could not resolve the IRS spectrum for TWA 5AB and therefore cannot conclude on the presence/absence of a disk around the brown dwarf TWA 5B. The age of the TWA ($\sim$10 Myr) makes it interesting for the study of brown dwarf disks at relatively older ages. In younger clusters, Luhman et al. (2005) have reported disk fractions of 42\% $\pm$ 13\% and 50\% $\pm$ 17\% for the sub-stellar ($>$M6, {\it M} $\leq$ 0.08 $M_{\odot}$) members of IC 348 (2-3 Myr) and Chamaeleon I ($\sim$1 Myr), respectively, while the fractions for stellar (M0-M6, 0.7 $M_{\odot}$ $\geq$ {\it M} $\geq$ 0.1 $M_{\odot}$) members are 33\% $\pm$ 4\% and 45\% $\pm$ 7\%. These fractions are based on observations made in the {\it Spitzer}/IRAC bands. In their IRAC and MIPS survey of brown dwarfs in Taurus (1-2 Myr), Guieu et al. (2007) have found a 48\% $\pm$ 14\% disk fraction, which is similar to that obtained by Hartmann et al. (2005) for T Tauri stars in Taurus. At $\sim$5 Myr, Carpenter et al. (2006) have reported a 19\% $\pm$ 5\% fraction for K0-M5 stars in Upper Scorpius, all of which show excess emission at 8 and 16 $\micron$, but none significant at shorter wavelengths, indicating a lack of warm dust in the inner disk ($\la$0.1 AU) region. In comparison, Bouy et al. (2007) have reported a much higher 50\% $\pm$ 16\% fraction for brown dwarfs (spectral type later than M6) in Upper Scorpius, half of which show clear excess at 8 $\micron$. In the same star-forming region, Scholz et al. (2007) have conducted a {\it Spitzer} survey of 35 brown dwarfs, and have reported a fraction of 37\%$\pm$9\%. These authors also report the presence of large inner holes ($>$ 5 AU), indicating an inner disk dissipation time scale of $\la$$10^{5}$ yr. At $\sim$10 Myr, the disk fraction for the stellar members of TWA is found to be 24\% (6/25), compared to 60\% (3/5) for the brown dwarfs. If we consider only the M0-M6 stars, then the fraction drops down to $\sim$16\% (3/18). A comparison with younger clusters then indicates that by the age of $\sim$10 Myr, the disk fraction for brown dwarfs has not decreased, whereas it drops by a factor of $\sim$2 for the higher mass stars. This indicates longer disk decay timescales for brown dwarfs compared to higher mass stars, consistent with the conclusions made by Bouy et al. (2007) and Scholz et al. (2007) on a mass-dependent disk lifetime. The low accretion rates in brown dwarfs would imply that their disks should live longer (Natta et al. 2004). The viscous scaling time for brown dwarfs are expected to be of the order of $10^{6}$ yr, longer than the typical $10^{4}$-$10^{5}$ yr for T Tauri stars (Alexander \& Armitage 2006). If disks persist for a longer time then the planet formation timescale might also be longer around brown dwarfs. Payne \& Lodato (2007) have investigated the potential for Earth-like planet formation around sub-stellar objects. Their simulations indicate that a rocky planet with an icy constitution and a mass as large as $\sim$0.1$M_{\earth}$ can form around a brown dwarf at a semi-major axis as small as $\sim$0.2 AU, for an average brown dwarf disk mass of 1.5 $10^{-4} M_{\sun}$ and an outer disk radius of 100 AU. However, the growth of such planetary cores is at a much slower rate around brown dwarfs compared to solar-type stars. Both 2M1207 and SSSPM 1102 show flat features due to highly processed grains. Best-fit disk models require large ($>$50 $\micron$) sized grains in the disk midplane, indicative of grain growth and dust settling. These are the processes that may lead to planet formation. On the other hand, rapid planet formation within $\sim$10 Myr may be responsible for the inner disk clearing observed in 2M1139. It might also be worth noting that the low-density environment of TWA might allow more brown dwarf disks to survive at early times. Though that would not explain the differences from stellar disks. A disk fraction of 60\% obtained here is considerably higher than the ones reported for the TWA stellar members, or for brown dwarfs in younger clusters. More sub-stellar members of TWA may remain to be identified, and their observations, along with brown dwarfs at older ages, will be valuable in determining if disks indeed persist for a longer time around sub-stellar objects compared to higher mass stars.

\acknowledgments
We wish to thank Barbara Whitney for helpful comments and suggestions. Support for this work was provided by NASA Research Grant $\#$ NNG06GJ03G. This work is based in part on observations made with the {\it Spitzer Space Telescope}, which is operated by the Jet Propulsion Laboratory, California Institute of Technology under a contract with NASA. Support for this work was provided by NASA through an award issued by JPL/Caltech. This work has made use of the SIMBAD database.

\end{document}